\documentclass{emulateapj}

\usepackage{apjfonts}
\usepackage{amsmath}
\usepackage{amssymb}
\usepackage{natbib}
\usepackage{xspace}
\usepackage{graphicx}
\usepackage{rotate} 

\newcommand{\fu}{4U~1907+09\xspace}

\newcommand{\Msun}{\ensuremath{M_\odot}\xspace}
\newcommand{\Lsun}{\ensuremath{\mbox{L}_\odot}\xspace}
\newcommand{\NH}{$N_{\text{H}}$~}

\shorttitle{4U 1907+09}
\shortauthors{Rivers et al.}

\begin{document}

\title{A Comprehensive Spectral Analysis of the  X-Ray Pulsar 4U~1907+09 from Two Observations with the {\it Suzaku} X-Ray Observatory} 

\author{Elizabeth~Rivers\altaffilmark{1}, Alex~Markowitz\altaffilmark{1}, Katja~Pottschmidt\altaffilmark{3}\altaffilmark{4}, Stefanie~Roth\altaffilmark{2}, Laura~Barrag\'an\altaffilmark{2}, Felix~F\"urst\altaffilmark{2}, Slawomir~Suchy\altaffilmark{1}, Ingo~Kreykenbohm\altaffilmark{2}, J\"orn~Wilms\altaffilmark{2}, Richard~Rothschild\altaffilmark{1}}

\altaffiltext{1}{University of California, San Diego, Center for
  Astrophysics and Space Sciences, 9500 Gilman Dr., La Jolla, CA
  92093-0424, USA} 
\email{erivers@ucsd.edu}
\altaffiltext{2}{Dr.\ Karl-Remeis-Sternwarte and Erlangen Centre for Astroparticle
  Physics, Sternwartstr.~7, 96049 Bamberg, Germany}
\altaffiltext{3}{CRESST and NASA Goddard Space Flight Center, Astrophysics Science
  Division, Code 661, Greenbelt, MD 20771, USA}
\altaffiltext{4}{Center for Space Science and Technology, University of Maryland
Baltimore County, 1000 Hilltop Circle, Baltimore, MD 21250, USA}

\begin{abstract}

We present results from two observations of the wind-accreting X-ray pulsar 4U~1907+09 using the 
\textsl{Suzaku} observatory.  The broadband time-averaged spectrum allows us to examine the continuum 
emission of the source and the cyclotron resonance scattering feature 
at $\sim$19\,keV.  Additionally, using the narrow CCD response of \textsl{Suzaku} near 6\,keV
allows us to study in detail the Fe K bandpass and to quantify the Fe K$\beta$ line for this source
for the first time.  The source is absorbed by fully-covering 
material along the line of sight with a column density of $N_{\rm H} \sim 2 \times 10^{22}$ cm$^{-2}$,
consistent with a wind accreting geometry, and a high Fe abundance ($\sim 3-4\,\times$ solar).  
Time and phase-resolved analyses allow us to study variations in the source spectrum.  
In particular, dips found in the 2006 observation which are consistent with earlier observations 
occur in the hard X-ray bandpass, implying a variation of the whole continuum rather than occultation 
by intervening  material, while a dip near the end of the 2007 observation occurs mainly in the lower
energies implying an increase in \NH along the line of sight, perhaps indicating clumpiness in the 
stellar wind.

\end{abstract}

\keywords{X-rays: stars --- X-rays: binaries --- stars: pulsars:
  individual (4U~1907+09) --- stars: magnetic fields}

\section{Introduction}

Discovered by Giacconi et al. (1971), the X-ray pulsar 4U~1907+09 is a wind-accreting 
high mass X-ray binary (HMXB) with a highly reddened companion star of magnitude 
$m_{\rm bol}=16.37$ ${\rm mag}$, luminosity $L_{\rm bol}=5\times 10^{5}$ \Lsun and a mass 
loss rate of $\dot{M}=7 \times 10^{-6}~\Msun~{\rm yr}^{-1}$ (Cox et al.\ 2005).
The neutron star is in an eccentric orbit with a period of 8.3753\,d.  
Orbital parameters can be found in in't Zand et al.\ (1998).

From 1983, when Makishima et al.\ (1984) found the source to have pulsations, through 
1998 the pulse period had a steady spin down rate of $\dot{P}_{\rm pulse}= +0.225{\rm \,s\,yr^{-1}}$, 
going from 437.5\,s to 440.3\,s (in 't Zand et al. 1998).  Baykal et al.\ (2006) reported a 
deviation from the spin down rate, measuring a much lower value between 1998 and 2003, 
which in 2002 was half the long-term value ($\sim0.115{\rm \,s\,yr^{-1}}$).  
Using \textsl{INTEGRAL} data, Fritz et al.\ (2007) demonstrated that between 2004 and 2005 
a complete torque reversal occurred with the maximum period of $\sim$441.3\,s being reached 
in April 2004 followed by spin up behavior with $\dot{P}_\textrm{pulse}= -0.158\,\pm\,0.007{\rm \,s\,yr^{-1}}$.  
\.Inam et al.\ (2009), using \textsl{RXTE}-PCA data between 2007 and 2008, reported a second torque 
reversal and a new spin down rate of $+0.220{\rm \,s\,yr^{-1}}$, consistent with the rate before 1998.

Using data from \textsl{Ginga} observations of 4U~1907+09 Makishima et al. (1992, 1999) 
reported a cyclotron resonance scattering feature (CRSF) at $\sim$19\,keV.  CRSFs or ``cyclotron lines'' 
are generated by photons scattering on electrons in the strong magnetic field near the surface of a neutron star.  
Photons from the accretion column above the magnetic poles that are near the energies that separate the Landau 
levels of the electrons are scattered out of the line of sight, the effect resembling an absorption line
(see, e.g., Nagel 1981; Sch\"onherr et al. 2007 for details).  
These features can thus be used to directly measure the magnetic field 
strength of the pulsar.  Using \textsl{BeppoSAX} observations Cusumano et al.\ (1998) later 
confirmed the CRSF for 4U~1907+09 at $\sim$19\,keV and calculated a strong surface magnetic field of 
$2.1 \times 10^{12}$\,G.  Cusumano et al.\ also reported a harmonic at $\sim$39 keV and an iron emission 
line near 6.4\,keV with an equivalent width of about 60\,eV.

4U~1907+09 was discovered to have an orbital period of 8.38 days by Marshall \& Rickets 
(1980) and to display periodic flaring behavior over the orbit as well as occasional dipping 
behavior.  The flaring was originally believed to be due to a Be type companion star, where 
the flares corresponded to the neutron star passing through the companion's outer atmosphere.  
Van Kerkwijk et al.\ (1989), however, optically identified the companion star 4U~1907+097 
as a supergiant, contradicting the earlier assumption.  This was later confirmed by Cox et al.\ (2005) 
who identified the companion as a type O8/O9 supergiant with a dense stellar wind, leading to 
the current belief that the pulsar is entirely wind-accreting.

The dipping behavior has been observed, notably by the \textsl{Rossi X-Ray Timing Explorer} (\textsl{RXTE}), 
as reductions in intensity for periods on the scale of an hour, occasionally dropping completely below 
detection levels (in't Zand et al.\ 1997).  Due to the absence of strong variations in the column 
density, the lack of correlation with the orbital phase, and various spectral features, in't Zand et al.\ 
have suggested that these dips are due to the cessation of mass accretion rather than occultation of intervening materal. 

In this paper we present data from the Japanese X-ray observatory \textsl{Suzaku} which observed 
the source twice, in 2006 and 2007.  In \S 2 we describe the instruments and the data reduction 
process.  In \S 3 we describe our methods and results for time-averaged spectral fitting, 
including details of the 6.4\,keV iron emission line and the CRSF at 19\,keV.  \S 4 shows the 
results of pulse-period determination and phase-resolved spectral fitting.  In \S 5 we detail 
time-resolved spectral fitting in order to explore the dipping behavior seen in both observations 
and the flaring seen in the 2007 observation.  Sections \S 6 and \S 7 are devoted to a detailed discussion
and a summary of our conclusions respectively.


\section{Data Reduction and Analysis}\label{sec:analysis}

\textsl{Suzaku} observed 4U~1907+09 with the X-ray Imaging Spectrometer (XIS; Mitsuda et al.\ 2007) and the 
Hard X-ray Detector (HXD; Takahashi et al.\ 2007) beginning 2006 
May 2 at 06:11 UT (MJD 53857.3) for 123\,ks and again 2007 April 19 starting at 10:03 UT 
(MJD 54209.4) for 158\,ks.  The observation ID's are 401057010 and 402067010 respectively.  
The HXD gathered data with both its detectors, the PIN diodes 
and the GSO scintillators, however we did not use the GSO data because of the faintness of 
the source in the GSO band relative to the non-X-ray background.  The XIS has 4 CCDs: 0--3 with 0, 2 
and 3 front-illuminated and XIS1 back-illuminated. In November of 2006 XIS2 was 
damaged and taken off-line, therefore data in the 2007 observation 
were taken with only the remaining three XISs.  Two corners of each XIS CCD have an $^{55}$Fe 
calibration source which can be used to calibrate the gain and test the spectral resolution 
of data taken using this instrument (See the \textsl{Suzaku} Data Reduction Guide 
\footnote{http://heasarc.gsfc.nasa.gov/docs/suzaku/analysis/abc/abc.html} for details).

During the 2006 observation the XIS was set to $1/4-$window mode to avoid saturating the 
CCD should the flux during the source's flares have proven to be too high.  However, this 
did not prove to be the case and for the 2007 observation the mode was switched to full-window. 
The pointing was also changed from XIS nominal pointing in 2006 to HXD nominal, both to reduce 
the flux in the XIS by a small fraction (10\%) and to increase the sensitivity above 12\,keV 
for analysis of the CRSF.

Data were processed with version 2.0.6.13 of the \textsl{Suzaku} pipeline and screened to 
exclude data within 436\,s of passing through the South Atlantic Anomaly and within Earth 
elevation angles of less than 5$^{\circ}$ or Bright Earth angles of less than 20$^{\circ}$.  
All extractions were done using HEASOFT v.6.6.2.


\subsection{XIS Reduction}

After screening, the good exposure time per XIS was 58.4\,ks for 2006 and 80.6\,ks for 2007.
The XIS CCDs were in the 3 $\times$ 3 and 5 $\times$ 5 editing modes which were cleaned and added 
to create image files for each XIS.  Data from each CCD were analyzed separately.  From these 
we extraced source and background lightcurves and spectra and used XISRMFGEN and XISSIMARFGEN 
to create the response matrix (RMF) and ancillary response (ARF) files. There was no 
charge injection in either observation.  Channels were grouped by 5 channels below 1.5\,keV 
and 10 channels from 1.5--10\,keV.~~Data below 1\,keV and above 10\,keV were ignored for all XIS 
in both observations, due to complete absorption below 1\,keV and instrument noise above 10\,keV. 
In addition, data between 1.8 and 2.7\,keV and between 1.65 and 2.3\,keV were ignored for the 2006 
and 2007 observations respectively, due to poorly understood calibration for the Si K complex and 
Au M edge arising from the detector and mirror system.

Fitting the $^{55}$Fe calibration sources for 2007 in XSPEC v.11.3.2ag (Arnaud et al.\ 1996) with a model comprised of 
three Gaussian components (Mn K$\alpha_{1}$, K$\alpha_{2}$ and K$\beta$) yielded the following 
results for the Mn K$\alpha_{1}$ line energy (expected value of 5.899\,keV):  5.829\,$\pm$\,0.004\,keV 
for XIS0, 5.886\,$\pm$\,0.006\,keV for XIS1 and 5.920\,$\pm$\,0.007\,keV for XIS3, showing that the 
energy calibration is good to 20\,eV in XIS1 and XIS3 and to 70\,eV in XIS0.  No calibration spectra 
were available for the 2006 data which was in $1/4-$window mode and therefore did not see the 
corners of the CCDs.


\subsection{PIN Reduction}

The Suzaku HXD/PIN is a non-imaging instrument with a 34$\arcmin$ square (FWHM) field of view.  
The HXD instrument team provide non-X-ray background (NXB) event files using the calibrated GSO 
data for the particle background monitor (``background D'' or ``tuned background'' with METHOD=LCFITDT).  
This yields instrument backgrounds with $\lesssim$1.5\% systematic uncertainty at the 1$\sigma$ 
level (Fukuzawa et al. 2009).  As suggested in the Suzaku ABC Guide, the Cosmic X-ray Background 
was simulated in XSPEC v.11 using the form of Boldt (1987).

Net spectra were extracted and deadtime-corrected for a net exposure time of 32.0\,ks for 2006 and 
64.9\,ks for 2007.  We excluded PIN data below 13\,keV due to thermal noise.  In both observations 
the source flux falls to levels well below a few percent of the background at $\sim$40\,keV for our 
time-averaged spectra and therefore we only use data between 13 and 40\,keV.  Channels were 
grouped with a minimum of 50 counts per bin to allow $\chi^2$ fitting. 


\section{Time-Averaged Spectral Analysis}\label{sec:specta}


\begin{figure}
\plotone{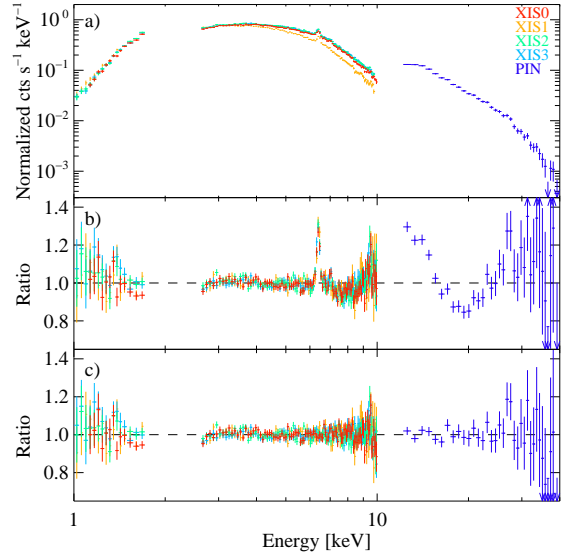}
\caption{Time-averaged spectral fitting for 2006.  Panel a) shows the data, all four XISs and the PIN.  Panel b) the data/model residuals 
for an absorbed continuum model with FDCut.  Panel c) shows the data/model residuals for our best-fit model 
including the iron lines, 10\,keV feature and the CRSF.}
   \label{fig1}
\end{figure}

\begin{figure}
\plotone{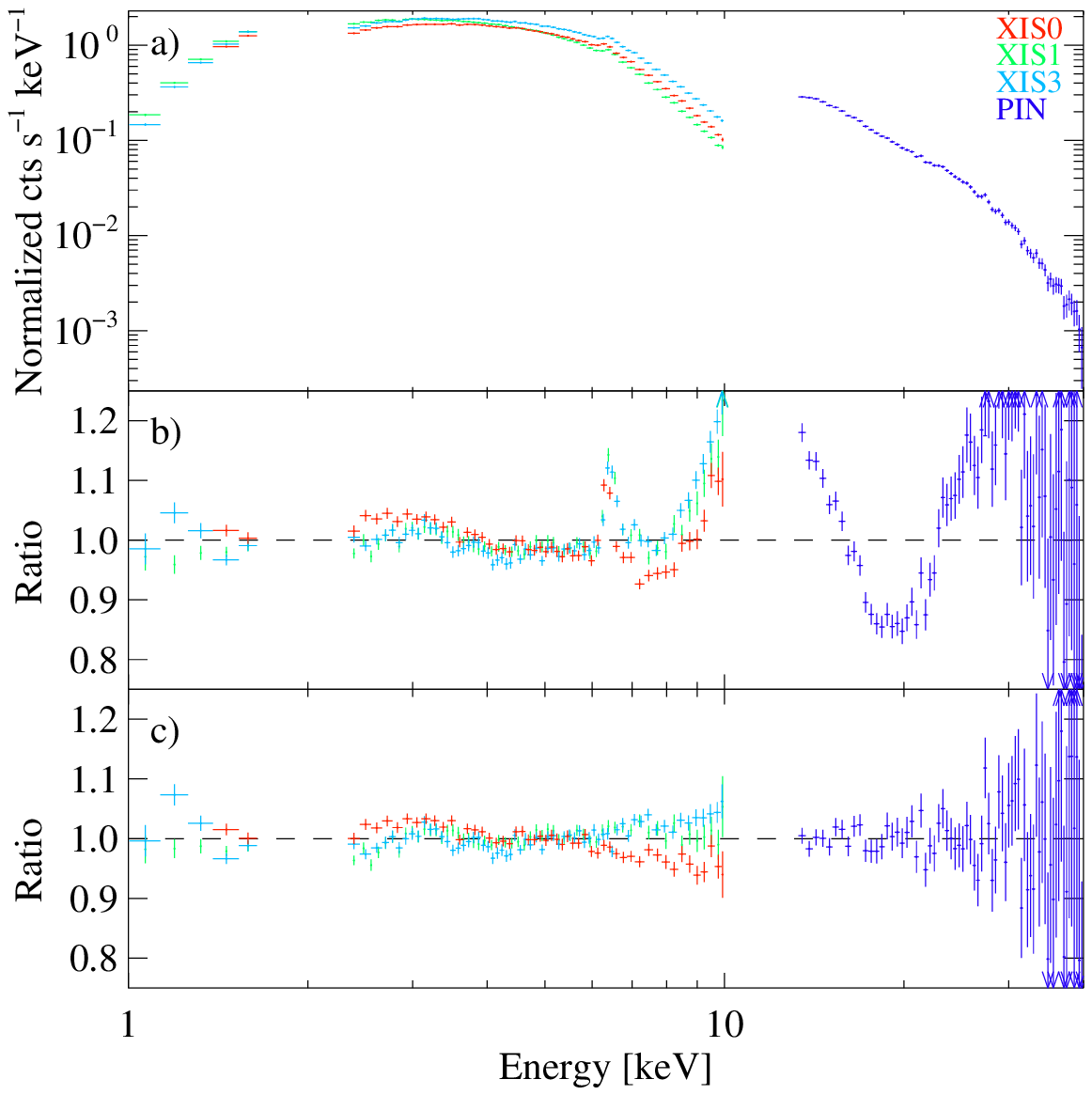}
\caption{Time-averaged spectral fitting for 2007. Panel a) shows the data, three XISs and the PIN.  Panel b) the data/model residuals 
for an absorbed continuum model with FDCut.  Panel c) shows the data/model residuals for our best-fit model 
including the iron lines, 10\,keV feature and the CRSF.}
   \label{fig2}
\end{figure}

\begin{figure}
\plotone{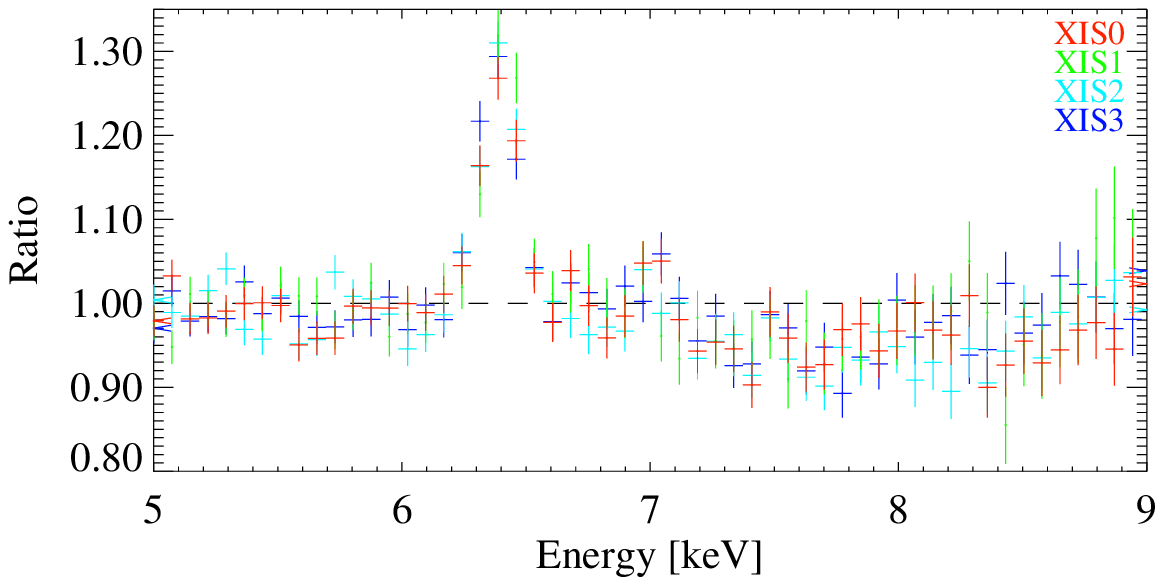}
\caption{Data/model residuals focusing on the iron K bandpass for 2006 when the continuum is modeled as in Fig.\ 1b.}
   \label{fig3}
\end{figure}

\begin{figure}
\plotone{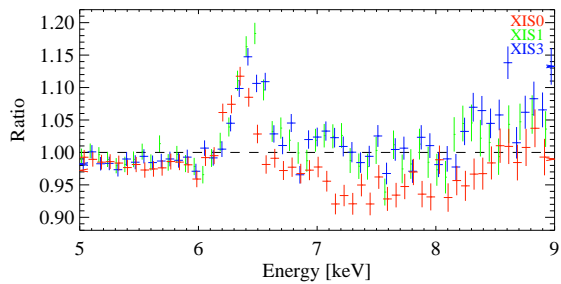}
\caption{Data/model residuals focusing on the iron K bandpass for 2007 when the continuum is modeled as in Fig.\ 2b.}
   \label{fig4}
\end{figure}


All spectral fitting in this paper used XSPEC v.11.3.2ag (Arnaud et al.\ 1996).
Unless otherwise stated, all uncertainties below correspond to 
$\Delta$$\chi^2$ = 2.71 (90$\%$ confidence level for one parameter).
For the time-averaged fits the normalization for XIS0 was fixed while the cross-normalizations were left free for 
XIS1, 2 \& 3 (2006) and for XIS1 \& 3 (2007); best-fit values of $\chi^2$/dof were usually close to unity.
The PIN/XIS0 cross-normalization was set to 1.16 (2006, XIS-nominal)
or 1.18 (2007, HXD-nominal pointing; Maeda et al.\ 2008).  
Allowing the cross-normalization to be free resulted in unrealistic values
due to a broad feature around 8--15\,keV.
We fit the XIS+PIN data in the range 1--40 keV.
The 2006 data are shown in Fig. 1a; the
2007 data are shown in Fig. 2a.

Our initial continuum model consisted of a power-law continuum 
emission component with a high-energy rollover, plus
full-covering absorption by a cold column of gas along the line of sight.
For both the 2006 and 2007 observations, we considered
three forms for the power-law continuum rollover:
a power-law with a smooth Fermi-Dirac cutoff ``FDCut'' (Tanaka 1986), 
a power-law with an abrupt exponential cutoff, ``HighECut'' (White, Swank \& Holt 1983),
and the ``NPEX'' model consisting of negative and positive power-laws 
with a common exponential cutoff (Mihara 1995). 
Qualitatively, all models gave similar overall results for each
observation.  For the purposes of this paper we will be using the FDCut model
to illustrate how spectral fitting proceeded and to fit the time- and phase-resolved spectra. 
We include the other two models as an alternate phenomenological descriptions of the continuum
and for ease of comparison with other observations.

For the absorber, we assumed the elemental abundances of Wilms, Allen \& McCray (2000)
and the photo-absorption cross sections of Verner et al.\ (1996); solar abundances were initially assumed.
Data/model ratios to this best-fit initial FDCut model
are shown in Figs.\ \ref{fig1}b and \ref{fig2}b for 2006 and 2007, respectively.
Initial values for $\chi^2$/dof were 2689.41/996 (2006) and 3820.61/784 (2007).

Figs.\ \ref{fig3} and \ref{fig4} show the data/model ratios in the Fe K bandpass.
Both spectra showed narrow, positive data/model residuals
at 6.4 keV, evidence for the Fe K$\alpha$ emission line.
We modeled this line using a Gaussian component. 
In the 2007 fit we allowed the line energy to be free for all three XIS's
based on the differences in gain observed in the calibration source.
This allowed for a much better fit of the K$\alpha$ line in the 2007 data.
It was not necessary to do this in the 2006 data.

Strong data/model ratios near $\sim$15--25 keV (Figs.\ \ref{fig1}b \& 
\ref{fig2}b) indicated the presence of the CRSF, which we modeled using
a Gaussian component in absorption. The actual shape of the CRSF is more complicated,
but for the purposes of fitting, a Gaussian is adequate to model the feature (Sch\"onherr et al., 2007).
In our best-fit model the Gaussian centroid energy was 18.6$^{+0.8}_{-0.7}$\,keV 
(2006) and 19.1$\pm$0.2\,keV (2007), consistent with previous results by 
Cusumano et al.\ (1998) and Makishima et al.\ (1992, 1999); note that 
Makishima et al.\ used a Lorentzian line profile while we and Cusumano et al.\
used a Gaussian, however the two profiles have similar energy centroids despite 
their different shapes.  We were not able to search for the second harmonic
at $\sim$39\,keV since the PIN spectrum becomes background dominated in that region.

We modeled positive ratios 
above 8 keV in the XIS and below $\sim$15 keV in the PIN with a broad 
emission feature peaking around 10--13\,keV.  The ``10 keV'' bump feature has been
observed  previously in this source by Mihara (1995) and Coburn et al.\ (2002) 
as well as in several other cyclotron line sources, e.g., with \textsl{RXTE} in
MXB~0656$-$072 (McBride 2006),\ Her~X-1, 4U~1626$-$67,
and 4U\,1538$-$52 (Coburn et al. 2002), and Cen~X-3 (Suchy et al. 2007); and with
\textsl{Ginga} in 4U~1538$-$52, and V\,0331$+$53 (Mihara 1995).
Though the physical nature of this feature is still not clear, it cannot stem from
calibration issues since it has been seen with multiple satellites.
It is believed to be part of the overall continuum and
including this feature as a broad Gaussian emission component in the model is required for a good fit.
$\chi^2$/dof fell from 1660.25/990 to 1125.39/987 (2006) and from 1743.56/776 to 1315.54/773 (2007).
The best-fit energy centroid for this component 
falls within or close to the gap in energy coverage between the XIS and PIN
and the effective areas of the XIS and PIN
are low in this region. The energy, width and intensity of this 
component as derived with \textsl{Suzaku} 
may therefore be affected by systematic uncertainties
which are difficult to quantify.

Next, we noticed narrow positive data/model ratios
near 7.1 keV, seen in all four XISes in 2006
and primarily in XIS1 and 3 in 2007. We identify this feature
as Fe K$\beta$ emission; detection of this line in \fu is reported here for the 
first time. Modeling it with a Gaussian component,
with the energy left as a free parameter and 
the width of the K$\beta$ line kept tied to that of the K$\alpha$ line,
$\chi^2$/dof fell to 1090.11/986 (2006) or 1290.08/770 (2007).
In our best-fit model, it was significant at the 99.994$\%$ (2006) or 99.998$\%$ (2007)
confidence level to include this component according to an $F$-test (which
formally cannot be used in this manner due to a boundary condition violation as per 
Protassov et al.\ (2002), but can give a rough approximation of the significance).
The best-fit value of the intensity was $13 \pm 5 \%$ 
that of the K$\alpha$ line, consistent with an origin in neutral
or at most moderately-ionized ($<$ \textsc{XIII}) species of Fe (Palmeri et al.\ 2003).
For the remainder of this paper, the K$\beta$ normalization is fixed to
13$\%$ that of the K$\alpha$ line.  

Finally, we noticed absorption-like data/model 
ratios between 7--8 keV. We fit this edge-like feature
by allowing the Fe abundance of the absorber
relative to solar, $Z_{\rm Fe}$, to vary. As shown in Table \ref{table1},
values of $Z_{\rm Fe}$ near 3--4 were typically obtained.
We thus obtained our best-fit model, with
$\chi^2$/dof falling to 1040.19/985 (2006) or 1239.02/766 (2007).
This edge is consistent with our values for the Fe K$\alpha$ and K$\beta$ lines, 
all of which have significant physical implications as will be discussed in \S6.3.

Figs.\ 1c and 2c show the data/model ratios for
our best-fit FDCut models, with all the above-mentioned components included.
Best-fit model parameters are listed in Table \ref{table1} for 
the FDCut, HighECut, and NPEX fits to the 2006 and 2007 spectra.
Flux values for the two observations were 1.5 and 2.8 $\,\times\,10^{-10}$\,ergs\,cm$^{-2}$\,s$^{-1}$.
The fits are robust to $\pm$2$\%$ variations in the PIN background level
(Fukazawa et al.\ 2009).


\begin{center}
\begin{deluxetable*}{llllllll}
   \tablecaption{Time-Averaged Spectral Fit Parameters \label{table1}}
   \tablecolumns{7}
\startdata
\hline
\hline\\[0.3mm]
Model                     &    FDCut        &  &  HighECut  &  &  NPEX  \\[1mm]
\hline\\[0.3mm]
Parameter                 &    2006  &   2007    &    2006   &   2007     &    2006    &   2007  \\[1mm]
\hline\\[0.3mm]
2--10\,keV Flux (ergs\,cm$^{-2}$\,s$^{-1}$)  & 1.5$\,\times\,10^{-10}$   & 3.8$\,\times\,10^{-10}$
                                           & - & - & - & -

\\[1mm]
\NH  ($10^{22}$ cm$^{-2}$)             & 2.9 $^{+0.1}_{-0.1}$        & 2.5 $^{+0.2}_{-0.2}$
                                      & 3.0 $^{+0.3}_{-0.2}$        & 2.3 $^{+0.2}_{-0.2}$
                                      & 2.7 $^{+0.1}_{-0.1}$        & 2.2 $^{+0.1}_{-0.1}$                  
\\[1mm]
Fe Abundance (rel. to Solar)          & 3.9 $^{+0.9}_{-0.7}$          &  3.0 $^{+1.0}_{-0.6}$ 
                                      & 2.8 $^{+0.6}_{-0.7}$          &  3.6 $^{+0.1}_{-0.1}$
                                      & 3.6 $^{+1.0}_{-0.8}$          &  3.3 $^{+0.6}_{-0.5}$
\\[1mm]
Photon Index                          & 1.08 $^{+0.02}_{-0.02}$      & 1.21 $^{+0.01}_{-0.01}$   
                                      & 1.09 $^{+0.03}_{-0.04}$      & 1.25 $^{+0.01}_{-0.01}$
                                      & -        &  -
\\[1mm]
Power Law Norm\tablenotemark{1}       & 2.36 $^{+0.07}_{-0.06}\,\times\,10^{-2}$~~~   & 5.07 $^{+0.04}_{-0.04}\,\times\,10^{-2}$   
                                      & 1.85 $^{+0.07}_{-0.06}\,\times\,10^{-2}$      & 5.07  $^{+0.04}_{-0.04}\,\times\,10^{-2}$
                                      & -         & -
\\[1mm]
Cutoff Energy (keV)                   & 13.6 $^{+1.3}_{-1.3}$          &  22.6 $^{+0.3}_{-0.3}$   
                                      & 20.4 $^{+2.9}_{-1.8}$          &  20.7 $^{+0.2}_{-0.2}$
                                      & -        &  -
\\[1mm]
Folding Energy (keV)                  & 9.5 $^{+0.4}_{-0.4}$          &  6.4 $^{+0.2}_{-0.2}$     
                                      & 6.6 $^{+1.0}_{-1.0}$          &  7.1 $^{+0.2}_{-0.2}$
                                      & -         &  -
\\[1mm]
NPEX $\alpha 1$                       & -    & -   & -   & -
                                      & 0.56 $^{+0.03}_{-0.02}$      & 0.65 $^{+0.01}_{-0.01}$                  
\\[1mm]
NPEX $\alpha 2$                       & -    & -   & -   & -
                                      & -2.00 (fixed)     & -2.00 (fixed)                 
\\[1mm]
NPEX Norm 1 ($10^{-2}$)                          & -    & -   & -   & -
                                      & 1.73 $^{+0.10}_{-0.07}$      & 4.73 $^{+0.08}_{-0.08}$                  
\\[1mm]
NPEX Norm 2 ($10^{-3}$)                           & -    & -   & -   & -
                                      & 2.5 $^{+1.0}_{-0.3}$      & 3.3 $^{+0.1}_{-0.1}$                  
\\[1mm]
NPEX Temperature (keV)                & -    & -   & -   & -
                                      & 4.9 $^{+0.1}_{-0.3}$      & 4.53 $^{+0.01}_{-0.01}$                  
\\[1mm]
CRSF Energy (keV)                    & 18.6 $^{+0.8}_{-0.7}$         &  19.3 $^{+0.2}_{-0.2}$  
                                      & 18 $^{+3}_{-1}$              &  19.9 $^{+0.7}_{-0.7}$
                                      & 18.4 $^{+1.0}_{-1.8}$         &  19.1 $^{+0.2}_{-0.2}$
\\[1mm]
CRSF $\sigma$ (keV)                     & 2.0 $^{+1.2}_{-0.9}$          &  2.8 $^{+0.3}_{-0.3}$  
                                      & 5.8 $^{+1.6}_{-0.8}$          &  3.2 $^{+0.4}_{-0.4}$
                                      & 3.6 $^{+1.7}_{-1.1}$          &  2.7 $^{+0.2}_{-0.2}$
\\[1mm]
CRSF Depth                           & 0.21 $^{+0.10}_{-0.06}$       &  0.41 $^{+0.03}_{-0.03}$ 
                                      & 1.0 $^{+0.1}_{-0.1}$          &  0.78 $^{+0.05}_{-0.08}$
                                      & 0.30 $^{+0.09}_{-0.05}$       &  0.47 $^{+0.01}_{-0.03}$
\\[1mm]
10 keV Bump Energy (keV)~~~        & 12.2 $^{+1.2}_{-0.2}$          &  11.4 $^{+0.1}_{-0.1}$
                                      & 12.8 $^{+0.3}_{-0.2}$          &  11.3 $^{+0.1}_{-0.1}$
                                      & 12.3 $^{+0.4}_{-0.4}$          &  11.2 $^{+0.1}_{-0.1}$
\\[1mm]
10 keV Bump $\sigma$ (keV)            & 1.5 $^{+0.1}_{-0.2}$       &  1.9 $^{+0.2}_{-0.2}$
                                      & 2.1 $^{+0.2}_{-0.2}$          &  1.4 $^{+0.2}_{-0.2}$
                                      & 1.5 $^{+0.4}_{-0.3}$          &  1.3 $^{+0.2}_{-0.2}$
\\[1mm]
10 keV Bump Intensity\tablenotemark{2}   & 3.2 $^{+0.5}_{-0.4}\,\times 10^{-3}$  & 3.8 $^{+0.3}_{-0.3}\,\times 10^{-3}$  
                                      & 5.7 $^{+1.8}_{-1.5}\,\times 10^{-3}$    & 2.5 $^{+0.2}_{-0.2}\,\times 10^{-3}$
                                      & 1.8 $^{+1.8}_{-0.5}\,\times 10^{-3}$    & 2.3 $^{+0.3}_{-0.3}\,\times 10^{-3}$
\\[1mm]
Fe K$\alpha$ Line Energy (keV)        & 6.394 $^{+0.005}_{-0.005}$       &  6.41 $^{+0.02}_{-0.02}$
                                      & 6.394 $^{+0.005}_{-0.005}$       &  6.43 $^{+0.02}_{-0.02}$
                                      & 6.394 $^{+0.005}_{-0.005}$       &  6.43 $^{+0.02}_{-0.02}$
\\[1mm]
Fe K$\alpha$ Line $\sigma$ (eV)       & < 35                           &  68 $^{+9}_{-21}$ 
                                      & 30 $^{+10}_{-16}$               &  66 $^{+14}_{-16}$
                                      & 28 $^{+16}_{-10}$               &  71 $^{+15}_{-15}$
\\[1mm]
Fe K$\alpha$ Line Intensity\tablenotemark{2}     
                                      & 2.0 $^{+0.1}_{-0.1}\,\times\,10^{-4}$~~~   &  2.5 $^{+0.2}_{-0.2}\,\times\,10^{-4}$  
                                      & 1.6 $^{+0.1}_{-0.1}\,\times\,10^{-4}$     &  2.3 $^{+0.2}_{-0.2}\,\times\,10^{-4}$
                                      & 1.4 $^{+0.1}_{-0.1}\,\times\,10^{-4}$    &  2.4 $^{+0.2}_{-0.2}\,\times\,10^{-4}$
\\[1mm]
Fe K$\alpha$ Line Equivalent Width (eV)      & 62 $\pm 3$            &  47 $\pm 4$
                                             & 67 $\pm 4$            &  46 $\pm 4$
                                             & 65 $\pm 4$            &  48 $\pm 4$
\\[1mm]
Fe K$\beta$ Line Energy (keV)         & 7.03 $^{+0.04}_{-0.04}$         &  7.0 $^{+0.1}_{-0.1}$ 
                                      & 7.02 $^{+0.04}_{-0.04}$         &  7.1 $^{+0.1}_{-0.1}$
                                      & 7.02 $^{+0.04}_{-0.04}$         &  7.0 $^{+0.1}_{-0.1}$
\\[1mm]
$\chi^2$/dof                          & 1040.19/985         &  1239.02/766
                                      & 1036.78/985         &  1248.23/766
                                      & 1018.14/985         &  1213.07/766
\\[1mm]
Reduced $\chi ^{2}$                   &  1.06         &  1.62
                                      &  1.05         &  1.63
                                      &  1.03         &  1.59
\tablenotetext{1}{Units are (ph\,keV$^{-1}$\,cm$^{-2}$\,s$^{-1}$ at 1\,keV)}
\tablenotetext{2}{Units are (ph\,cm$^{-2}$\,s$^{-1}$)}
\enddata
\end{deluxetable*}
\end{center}


\section{Phase Resolved Spectroscopy}\label{sec:phase}


\begin{figure}
\plotone{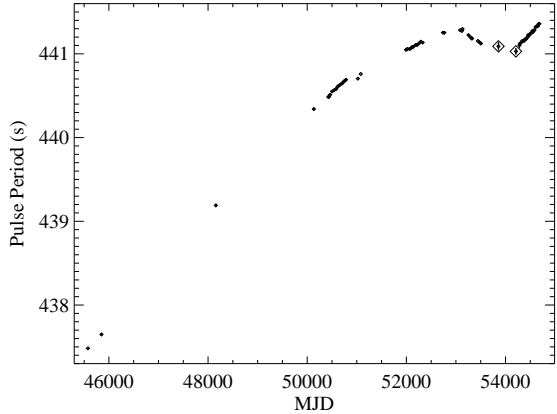}
\caption{Pulse period evolution from 1983 to 2008, using data from Baykal et al.\ (2001, 2006), Fritz et al.\ (2006), 
         \.Inam et al.\ (2009) and this work (indicated by diamonds).}
   \label{pulseev}
\end{figure}

\begin{figure}
\plotone{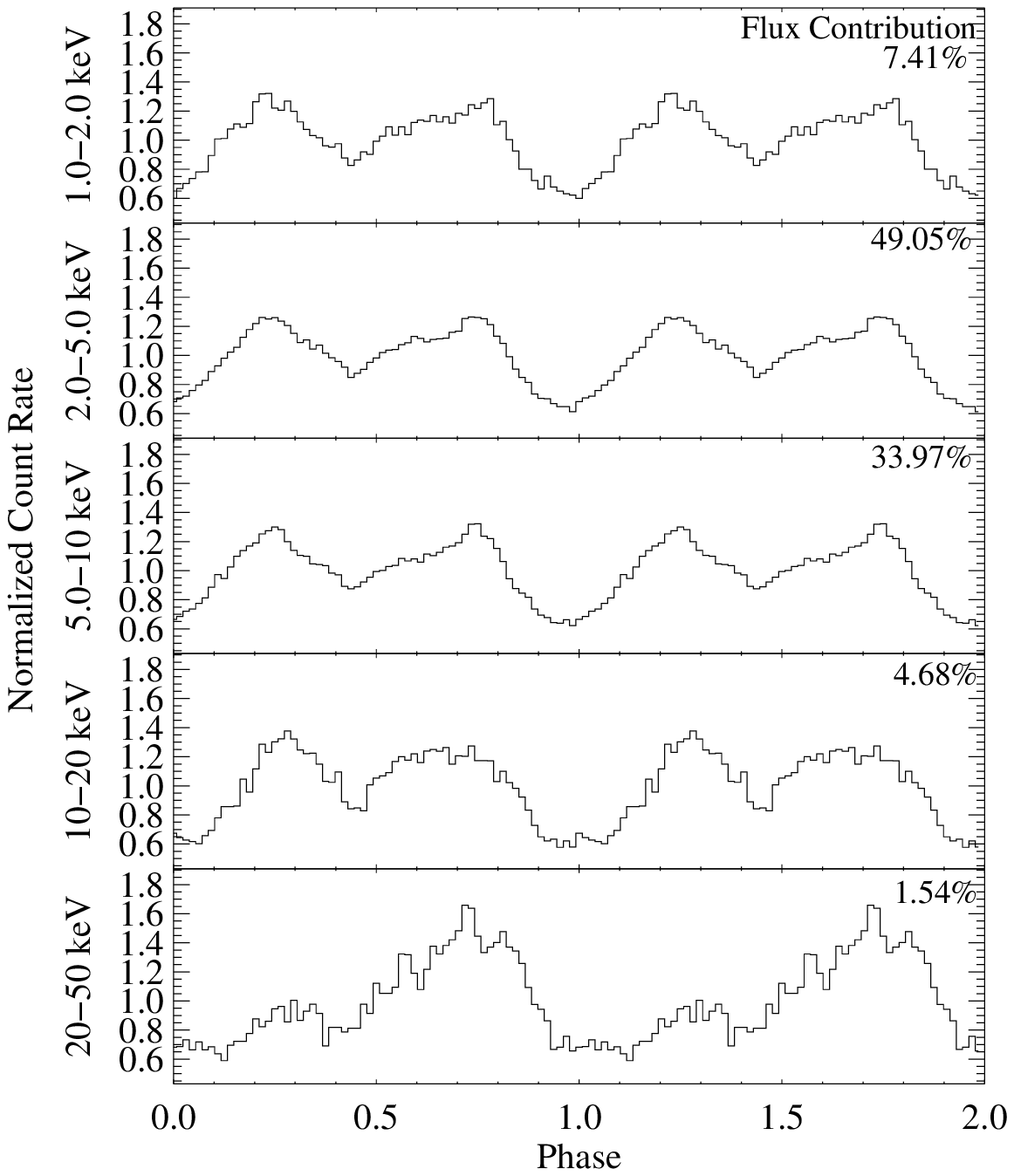}
\caption{Energy resolved pulse profiles for the co-added XIS and PIN data from 2006.  Flux contributions are listed for each energy band, representing the percentage of the total flux contributed by that energy range.}
   \label{pp06}
\end{figure}

\begin{figure}
\plotone{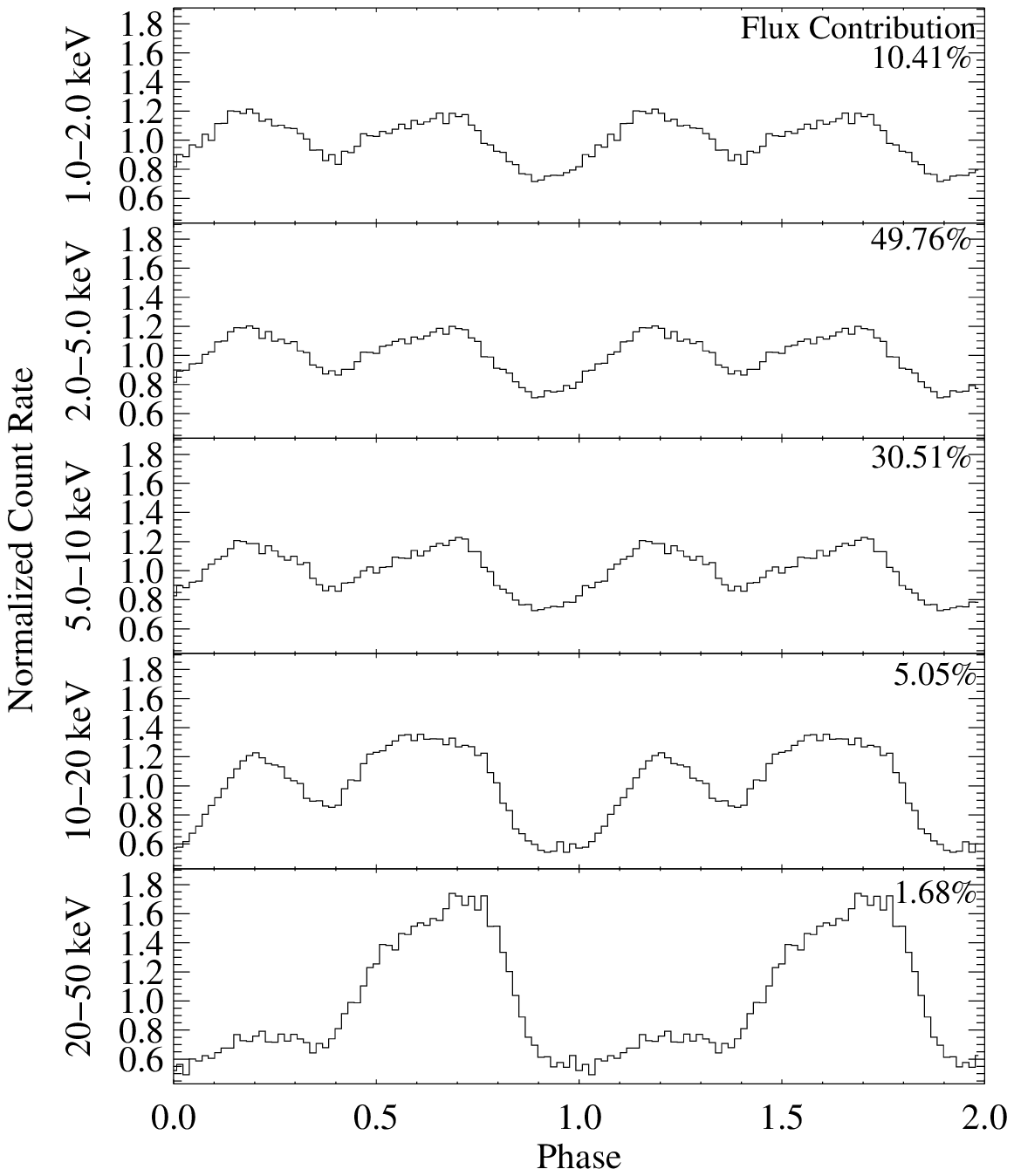}
\caption{Energy resolved pulse profiles for the co-added XIS and PIN data from 2007.  Flux contributions are listed for each energy band, representing the percentage of the total flux contributed by that energy range.}
   \label{pp07}
\end{figure}

\begin{figure}
\plotone{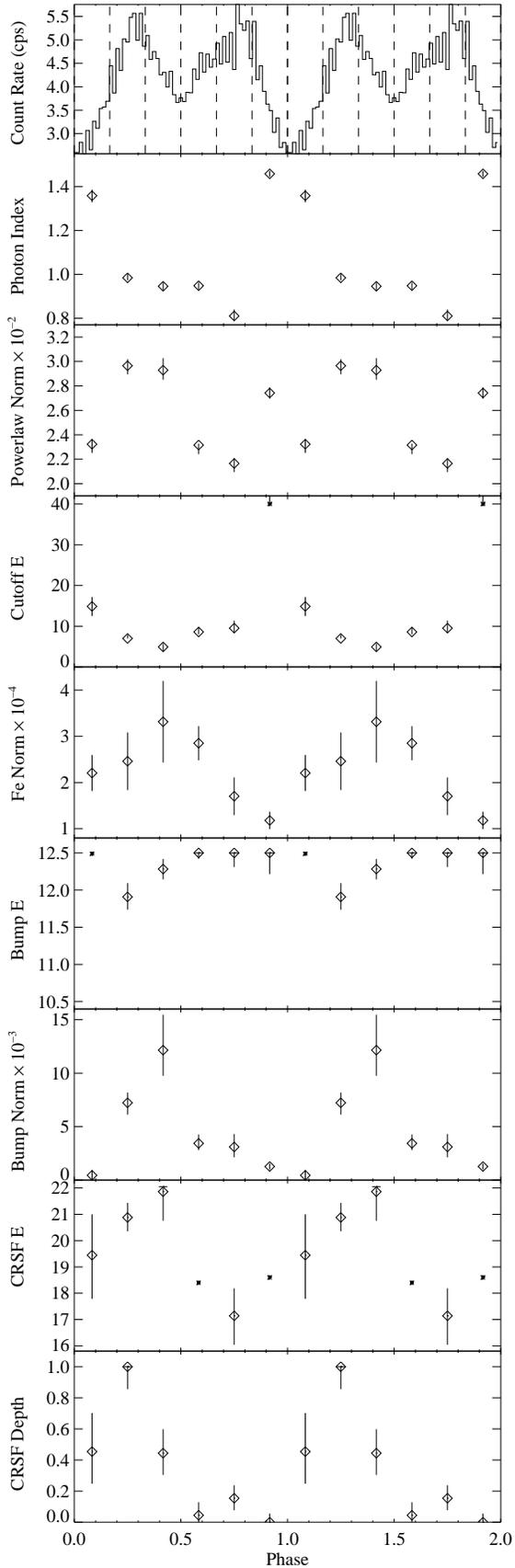}
\caption{Results for phase resolved spectral fitting for 2006.  Pulse profile is for XIS0.  For units see Table 2.
            A star symbol (*) indicates that the energy was unconstrained because the component was not needed for good fit. 
            The CRSF depth is capped at 1.0 and the ``10\,keV'' bump energy is capped at 12.5\,keV in order to avoid degeneracy 
            between these two components.}
   \label{phasefit06}
\end{figure}

\begin{figure}
\plotone{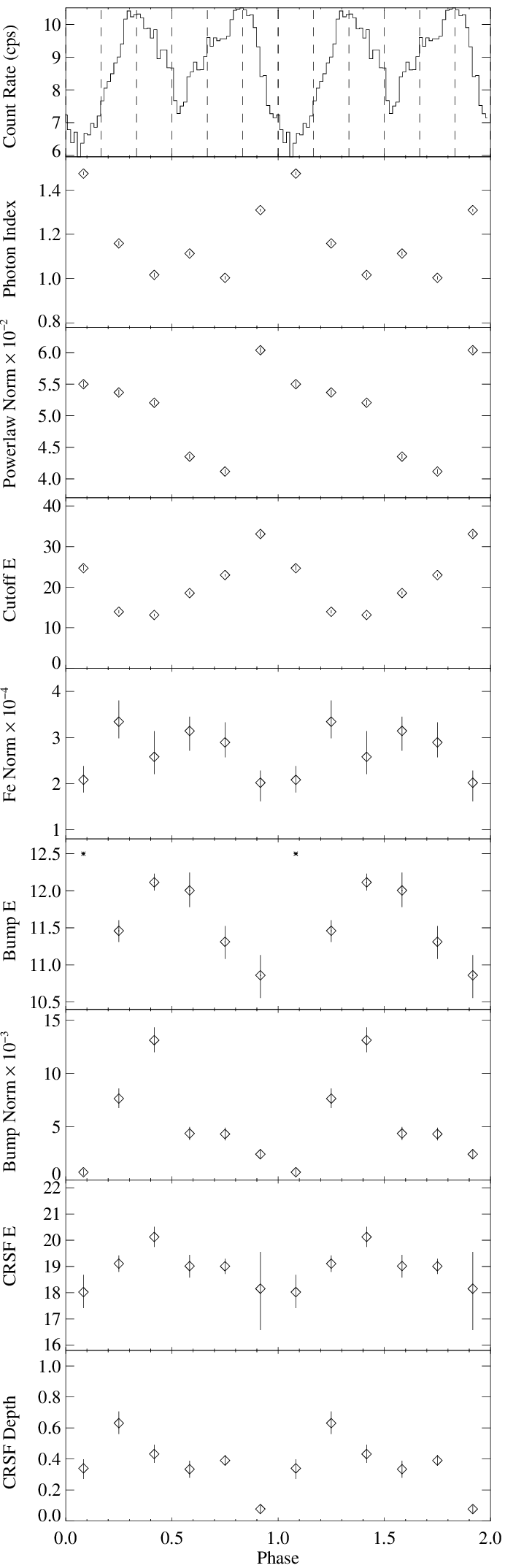}
\caption{Results for phase resolved spectral fitting for 2007.  Pulse profile is for XIS0.  For units see Table 2.
            A star symbol (*) indicates that the energy was unconstrained because the component was not needed for good fit.}
   \label{phasefit07}
\end{figure}


\subsection{Pulse Period Determination}\label{sec:puls}

To measure the pulse period we performed epoch folding (Leahy et al. 1983) on the PIN data for each observation, 
excepting the gaps due to the satellite orbit and only using the first $\sim$ 50\,ks for the 2006 observation to avoid the
pronounced dip beginning midway through that observation.  The 2006 observation pulse period was found to be 441.09$\pm 0.05$\,s while 
the 2007 observation pulse period was found to be 441.03$\pm 0.03$\,s.  Uncertainties are due to strong pulse-to-pulse variations 
as well as the brevity of our observations compared to the length of the pulse period.  Nevertheless, these numbers 
are consistent with the trends in the pulse period seen by both Fritz et al.\ (2006) and \.Inam et al.\ (2009).  The long-term 
evolution of the pulse period is shown in Fig.\ \ref{pulseev} with the two \textsl{Suzaku} observations falling in the midst of
the second torque reversal, restoring the source to its previous long-term spin-down rate (\.Inam et al.\ 2009).  
Physical implications of this event will be discussed in \S 6.

Using these periods we generated pulse profiles for each observation.  The profiles are shown in 
Figs.\ \ref{pp06} and \ref{pp07}, exhibiting a double-peak with the initial peak being much 
weaker in the higher energy bands.

\subsection{Phase Resolved Spectral Analysis}\label{sec:specph}

We divided the pulse profiles into 6 regions as shown in the first sections of Figs.\ \ref{phasefit06} 
(2006) and \ref{phasefit07} (2007) and extracted spectra for each region separately.  
Each bin had an average good exposure time per XIS of 9.5\,ks for 2006 and 13.2\,ks for 2007.
The number of bins was chosen as a trade-off between achieving good time or phase resolution and
having adequate signal/noise within each sub-spectrum.  We applied the best-fit time-averaged FDCut models,
allowing the following parameters to be free:  the power-law photon index ($\Gamma$), 
the power-law normalization, the cutoff energy, 
the CRSF energy and depth, the Fe line intensity and the 10 keV bump energy and depth.  
Other parameters were kept fixed in the fits at their time-averaged values;  for some, such as the e-folding energy,
there was no evidence of variability when these parameters were left free, 
thus we kept these parameters fixed to minimize any potential 
systematic effects on other free parameters.  The iron line energies and widths were also kept fixed since we do not expect 
them to vary with phase.  The results for selected model parameters
are plotted in Figs.\ \ref{phasefit06} (2006) and \ref{phasefit07} (2007).  
We caution, however, that with a low number of pulse bins, evidence for 
variation in these parameters as well as correlation with flux is not highly robust.

For both sets of data we find that the folding energy does not exhibit strong evidence 
of variation over the pulse phase. The cutoff energy and $\Gamma$ do seem to vary over the phase, 
with both parameters reaching high values in bins 0 and 5 and minimum values near bins $\sim$2--4.
The Fe line depth shows some variation over the pulse with minimum values in the largest dip. 
There also seem to be small variations in the CRSF parameters, all of which reach 
their largest values during the first pulse peak.  However, given the low number of counts above 10\,keV during the 
first peak and the influence of the parameters of the 10\,keV feature produce a lack of robustness in this result.
The fit results also suggest that the properties of the 10 keV feature may vary as a 
function of pulse phase, but as we cautioned earlier, there may be systematic effects 
hampering our ability to measure the properties of this feature accurately.


\section{Time Resolved Spectroscopy}\label{sec:time}

\begin{figure}
\plotone{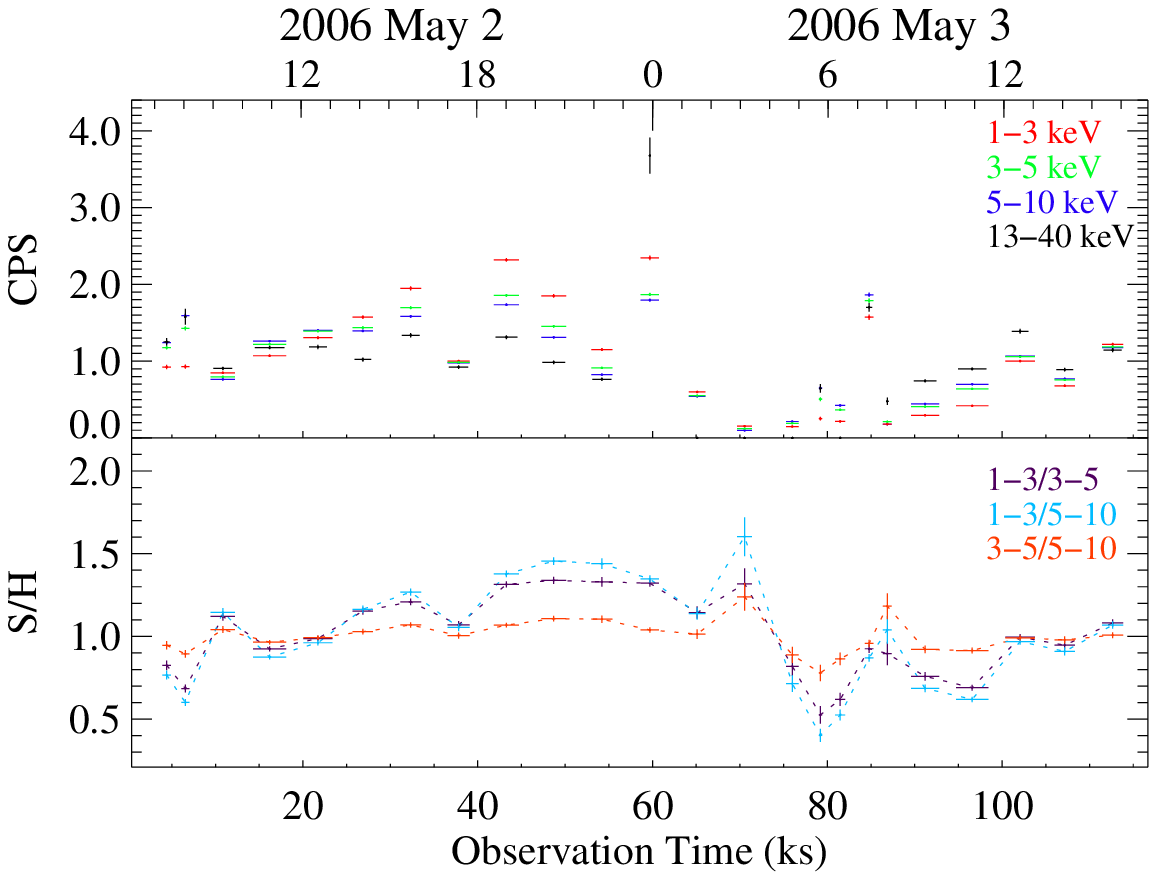}
\caption{Energy resolved, background subtracted, lightcurves and softness ratios from XIS0 for 2006.}
   \label{softness06}
\end{figure}

\begin{figure}
\plotone{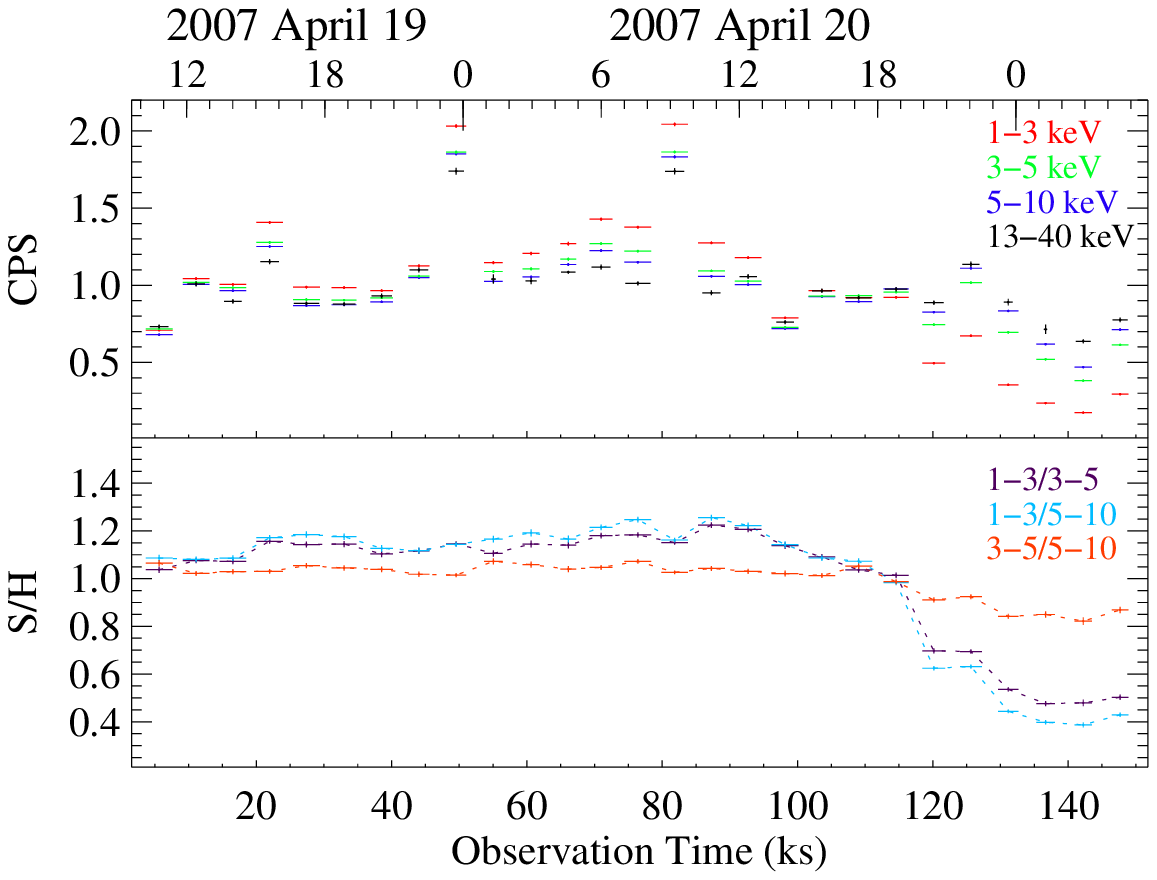}
\caption{Energy resolved, background subtracted, lightcurves and softness ratios from XIS0 for 2007.}
   \label{softness07}
\end{figure}

\begin{figure}
\plotone{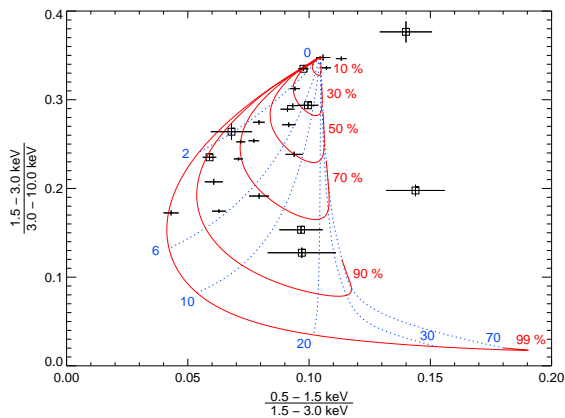}
\caption{Color-color diagram for the 2006 observation with a partial covering grid where solid lines are
covering fraction contours and dotted lines show constant $N_{\rm H}$. Boxed data points correspond to
the deep dip between $\sim$60\,ks and 100\,ks and do not fit on the partial covering grid.}
   \label{color06}
\end{figure}

\begin{figure}
\plotone{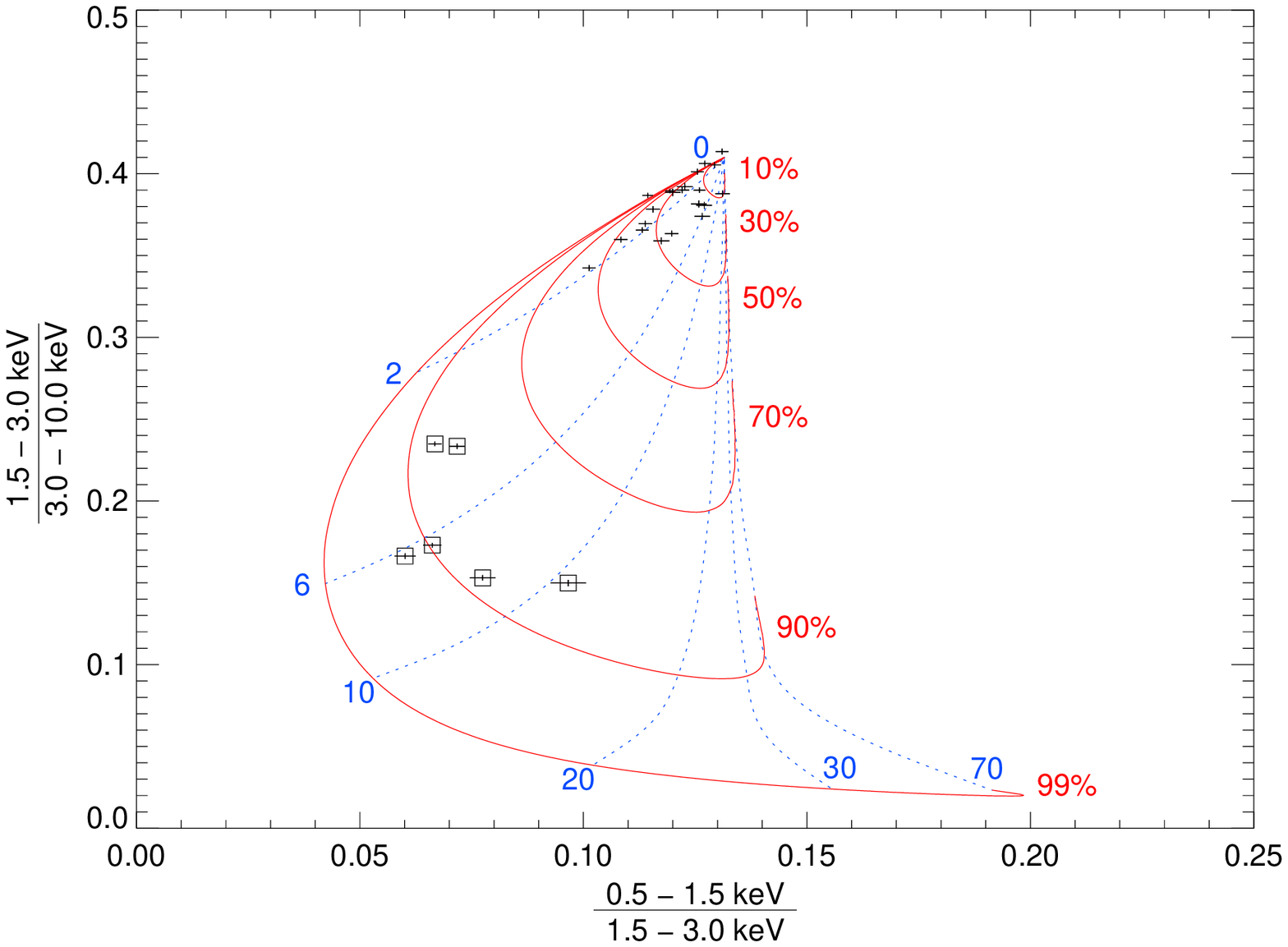}
\caption{Color-color diagram for the 2007 observation with a partial covering grid where solid lines are
covering fraction contours and dotted lines show constant $N_{\rm H}$. Boxed points correspond to the
dip in the last $\sim$40\,ks of the observation and are consistent with a partial coverer.}
   \label{color07}
\end{figure}

For a preliminary exploration of spectral variability as a function of time, we present count rate light curves binned
on the timescale of the satellite orbit in 
Figs.\ \ref{softness06} (2006) and \ref{softness07} (2007). Softness ratios are also shown, including (1--3)/(3--5)\,keV, 
sensitive to variations in $N_{\rm H}$, and (3--5)/(5--10)\,keV, sensitive to variations in $\Gamma$.  
Additionally, Figs.\ \ref{color06} and \ref{color07} show hardness ratios plotted against each other for each interval.
The observational data points in these color-color plots were obtained
adding data from XIS0 and XIS3, while the grids were created by
deriving colors from simulated spectra. The latter are based on the
spectral parameters of the point with the highest value of 1.5--3\,keV
over 3--10\,keV (excluding points from the dips and flares) and allowing
for an additional absorption component, partially covering the X-ray
source, i.e., the ``pcfabs'' component in XSPEC. Using the response
matrices for XIS0 and XIS3, hardness ratios for a grid of values of the
covering fraction and the additional \NH were calculated and then
interpolated to create the solid and dotted lines in Figs.\ \ref{color06} and \ref{color07}.


For the 2006 observation, 
the softness ratios vary over time, beginning and ending hard, while softening in the middle, suggesting that \NH is varying 
over the observation. The prolonged dip in 2006 between 60 ks and 100ks shows an interesting behavior, beginning very 
soft but then hardening quickly.  These points are shown in Fig. \ref{color06} as squares and clearly do not lie along the 
partial covering grid, indicating that this dip is not caused by a dramatic increase in $N_{\rm H}$.  
Together with the softness at the beginning of the dip, this implies a variation in $\Gamma$ 
and that the cause of the dip is not absorption but is primarily associated with the continuum component.
For the 2007 observation, the (1--3)/(3--5)\,keV softness ratio light curve exhibits a dip during the last $\sim$40\,ks of the
observation, suggesting an increase in $N_{\rm H}$, possibly caused by, for example, a relatively dense clump of material 
traversing the line of sight.  These dip points (indicated by squares in Fig.\ \ref{color07}) are consistent with a high density
partial covering material in the line of sight.
The (3--5)/(5--10)\,keV softness ratio light curve is roughly constant, suggesting 
that $\Gamma$ does not strongly vary, except possibly towards the final $\sim$20\,ks.

\begin{figure}
\plotone{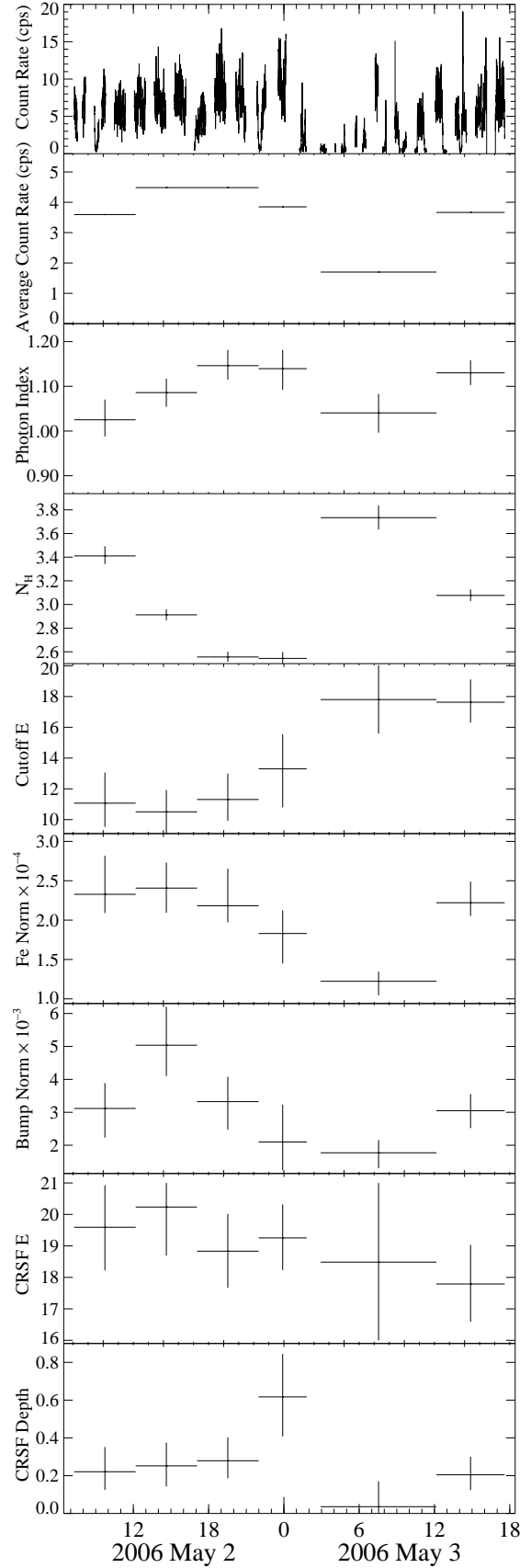}
\caption{Spectral fitting over time for 2006.  Light curve shown is for XIS0.  For units see Table 2.}
   \label{timefit06}
\end{figure}

\begin{figure}
\plotone{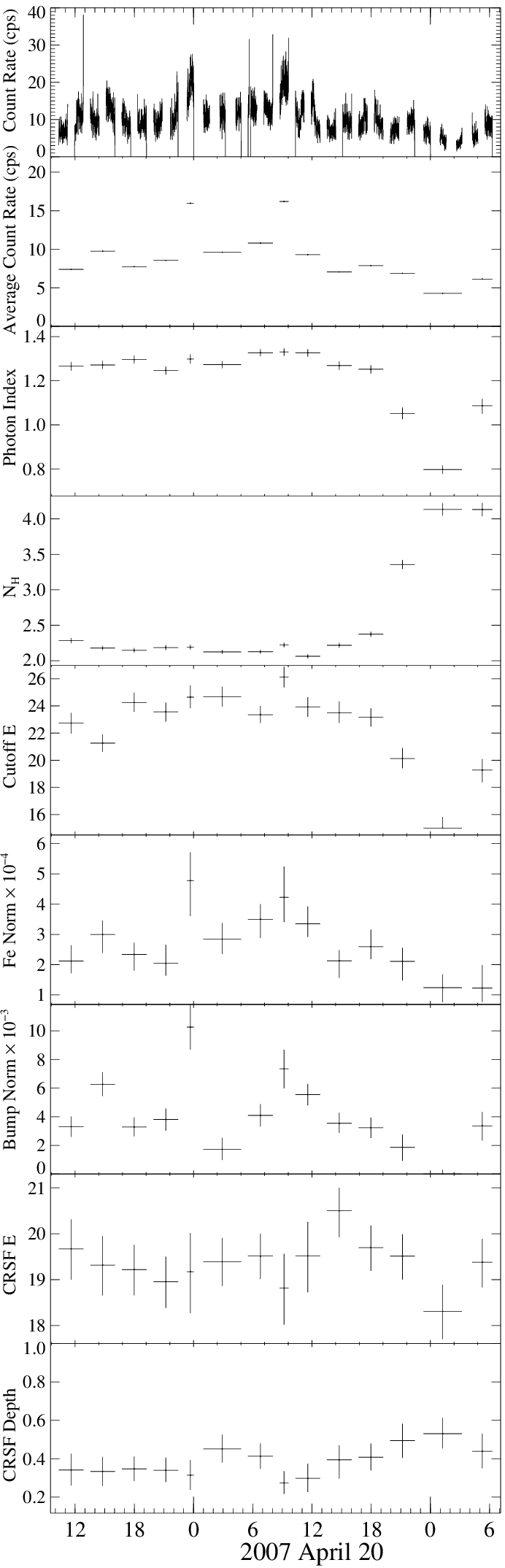}
\caption{Spectral fitting over time for 2007.  Light curve shown is for XIS0.  For units see Table 2.}
   \label{timefit07}
\end{figure}

We divided the 2006 and 2007 data into 6 and 14 bins respectively, trying to ensure 
enough counts in each bin to make spectral fitting possible, resulting in each bin having an 
average good exposure time of roughly 11\,ks (2006) or 5\,ks (2007).  The significant difference 
in binning for the two observations stems from the extended dip in 2006 (see the top panel of  
Fig.\ \ref{timefit06}) which caused the average flux to be $\sim$2.5 times fainter than 2007.
We applied the best-fit time-averaged FDCut models, keeping the iron line energies and widths, the
10\,keV feature energy and width, and the e-folding energy fixed at the time-averaged value.  
The light curves and selected model component parameters are 
plotted in Figs.\ \ref{timefit06} (2006) and \ref{timefit07} (2007).

In both observations we notice that the Fe line depth is strongly correlated with the flux.  Correlation
coefficients were found to be 0.86 for 2006 and 0.95 for 2007 with null-hypothese probabilities of 
97\% and $>$\,99.99\% respectively.  
The implications for the geometry of the Fe line-producing material around the source will be discussed in \S 6.  

For the 2006 observation we do see some variation in \NH and $\Gamma$, particularly during the dip
between 60 and 100\,ks.  2D contour plots of \NH versus $\Gamma$ calculated with the cutoff energy 
left as a free parameter for the lowest flux subspectrum show no strong correlation between 
the two parameters; additionally Monte Carlo simulations show a distribution 
in parameter values only on scales similar to the calculated error bars which are very 
small compared to the observed parameter variations.  Standard deviations for \NH and $\Gamma$ were
0.02\,cm$^{-2}$ and 0.015 respectively. In addition, we investigated possible correlations between $\Gamma$ 
and the cutoff energy in the low flux states but again 2D contour plots and Monte Carlo simulations
indicated that the degree of correlation was much smaller than the observed variation.

\begin{figure}
\plotone{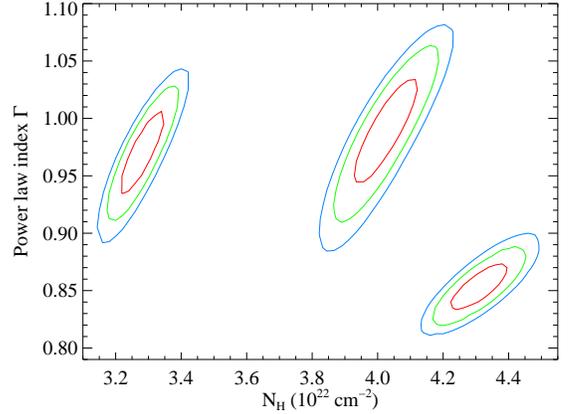}
\caption{Contour plots of \NH versus $\Gamma$ for the last three time bins of the 2007 time-resolved spectra.}
   \label{cont}
\end{figure}

For the 2007 observation there is a doubling in $N_{\rm H}$ during the final 2--3 spectra, from 
$\sim 2\,\times\,10^{22}$ to $\sim 4\,\times\,10^{22}$\,cm$^{-2}$.  Through most of the observation
$\Gamma$ is roughly constant, but then it too changes in the final 3 spectra, flattening 
significantly where the flux reaches its lowest values.  Contour plots of $\Gamma$ as a function of  $N_{\rm H}$
for the final three spectra yielded 
no evidence for any strong degeneracy between 
these two parameters from one time bin to the next (Fig.\ \ref{cont}).  
It is possible however, that the dip in $\Gamma$ may actually be a result of not using a partial covering model
for the absorber, since we see in Fig.\ \ref{color07} that the dip is consistent with partial covering.
The cutoff energy also decreased during this time, but we caution that there may be systematic 
effects associated with the lowest flux levels.  There is no strong evidence for variation 
in the parameters of the CRSF or the 10\,keV bump energy and width, however the 10\,keV bump normalization
appears to track the continuum flux, consistent with this feature being a part of the emission continuum
we do not yet fully understand.

\section{Discussion}\label{sec:disc}

\subsection{Properties of the CRSF}
 
Observations of CRSFs allow direct measurement of the neutron star's
magnetic fields, as the line energy $E_\text{cyc}\propto B/(1+z)$,
where $z$ is the gravitational redshift of the line-forming region.
Their properties (depths, profiles, etc.) depend on the pulsar X-ray
emission processes and the behavior of matter in strong magnetic
fields (Meszaros 1985). CRSFs thus carry a wealth of information
about the environment in which they are produced. 
For \fu we measure the energy centroid of the CRSF to be
consistent with previous measurements, e.g., that of
Cusumano et al.\ (1998) who calculated a magnetic field of $\sim2.1 \times 10^{12}$\,G
using a value of $(1 + z)^{-1}$ = 0.76.    
The depth and width are also consistent with previous measurements. 

In recent years it has been discovered that in most accreting neutron
stars the cyclotron line energy depends on the source flux
(Mihara 2004, 2007; Nakajima 2006). Prominent recent examples
for such a variability are V0332$+$53 and 4U~0115$+$63, where the
cyclotron line energy \emph{decreases} with flux
(Mowlavi 2006; Tsygankov 2006; Nakajima 2006), and Her~X-1, where the
line energy \emph{increases} with flux (Staubert et al.\ 2007). The most
prominent counter-example is A0535$+$262, where the line does not show
any flux dependence (Caballero et al.\ 2007). The line variability can be
understood by considering that for many accreting neutron stars, even
though the overall source luminosity is only at a few percent of
Eddington, the radiation field produced in the accretion column is
close to or above the \emph{local} Eddington limit. In this case, the
accretion stream is rapidly decelerated by radiation pressure from the
magnetic poles leading to a radiative shock, close to which the
cyclotron lines are formed (Becker \& Wolff 2005, 2007). The location
of the shock depends on the ram pressure of the accretion stream and
on the luminosity, $L$. For $L$ close to the local Eddington
luminosity, with increasing $L$ the shock moves away from the neutron
star, where the $B$-field is weaker. Thus the CRSF moves to lower
energies, as is indeed observed in most sources. For smaller $L$,
however, the ram pressure of the accreting material becomes important.
As shown by Staubert et al.\ (2007) for the case of Her~X-1, once $L$ is
below a critical luminosity the height of the column decreases as
$\dot{M}$ (and $L$) increases, because the accreted material
``squeezes'' the accretion mound. In this regime, therefore, the
cyclotron line energy increases with luminosity.
4U~1907$+$09, however, is more like A0535$+$262, with no observed
correlation between the CRSF energy and total flux. This could imply that
it is somewhere between the two regimes described above.

We see strong variations in the CRSF parameters over the pulse phase, as well
as in the continuum paramters, particularly $E_{\rm cutoff}$ and $\Gamma$.  These
clearly indicate that with a change in viewing angle we see different aspects
of the accretion column structure.  For details on phase-resolved pulsar spectra
see Coburn et al.\ (2002).


\subsection{The Pulse Period Evolution}

As seen in \S 4, \fu has an interesting and as it turns out, rather unique pulse period history.
Typical XRBs tend to maintain a trend of either spin-up or spin-down with short, intense
episodes of the contrary superimposed (Bildsten et al. 1997).  Examples of such behavior include
Her~X-1, Cen~X-3, Vela~X-1, and X~Per. These shorter episodes are often interpreted
to be due to short term variations in the mass transfer rate from the donor star 
which give rise to short-term torque fluctuations (Ghosh \& Lamb 1979).
Such changes are always accompanied by large magnitude changes in $L$.
The model used to describe this behavior is a magnetic coupling between the neutron 
star and its truncated accretion disk which transfers angular momentum to the neutron star,
the accreted matter producing a torque and giving rise to non-zero values of $\dot{P}$ 
(see, e.g., Ghosh, Pethick \& Lamb 1977, and references therein).

There are only three sources known to show a very different behavior: The
LMXB GX~1+4, the ultracompact LMXB 4U~1626$-$67, and \fu, all three of which
were historically characterized by nearly constant $\dot{P}$ over decade long
periods with no short-term fluctuations but rare dramatic reversals of $\dot{P}$
(Chakrabarty et al.\ 1997a,b; Bildsten et al. 1997; Fritz et al. 2006).
Monitoring of GX~1+4 by the \textsl{Fermi} Gamma-ray Burst Monitor show it continuing in its new trend 
\footnote{http://gammaray.msfc.nasa.gov/gbm/science/pulsars/},
however 4U~1626$-$67 has shown another torque reversal (Camaro-Arranz et al.\ 2009), 
returning to a spin up trend after 18 years of spin down.
Despite the difference of HMXB versus LMXB, \fu and 4U~1626$-$67 have shown very similar
pulse period behavior, both with $|P/\dot{P}|$$\sim$ several
1000\,years, neither showing any change in X-ray luminosity during reversals of $\dot{P}$,
and both displaying similar $P/\dot{P}$ ratios for their spin up and spin down periods.

With the simple magnetic torquing model outlined above, both the long
distinct episodes of constant $\dot{P}$ and the magnitude and
sign of $\dot{P}$ are difficult to explain together (Fritz et al. 2006, and references
therein).  As was observed in GX~1+4 (Chakrabarty et al.\ 1997a), one would expect dramatic changes 
in luminosity to accompany any torque reversal.  No such correlation has been seen
in either 4U~1626$-$67 or \fu.  Furthermore, it is difficult to understand
why the torquing in the spin up regime has a similar magnitude but
reversed sign than in the spin down regimes.

Recently, Perna, Bozzo \& Stella (2006) presented a new model in which torques can
change without having to invoke retrograde disks or fluctuations in
$\dot{M}$. This oblique rotator model can account for all observed properties of
the spin history of 4U~1626$-$67, including the large values of
$|P/\dot{P}|$ before and after the torque reversal and the virtually
unchanged luminosity of the source. However its prediction that such torque
reversals must be very rare is contradicted by the second observed
reversal of \fu in only a few years.  Further monitoring of the pulse period
will be needed as the evolution of the period of \fu may help to improve 
models of the accretion mechanism in XRBs.


\subsection{The Fe lines and the Absorbing Gas: Tracing the Accretion Flow}

We have studied the Fe K emission complex in detail, resolving the Fe K$\alpha$ line,
detecting Fe K$\beta$ for the first time in this object, and detecting a strong Fe K edge at 7.11 keV
which we model employing an Fe abundance 4 times the solar value in the absorbing material.

It is possible that the same material which absorbs the X-ray continuum below 2 keV along the line of sight
is responsible for transmitting the fluorescent emission lines.  We can assume for simplicity
that the absorbing gas is situated in an optically-thin, uniform-density,
spherical shell of material surrounding the central X-ray source.
We can thus use Eqn.\ 4 of Yaqoob et al.\ (2001; based on Krolik \& Kallman 1987),
which yields an estimate for the equivalent width, EW$_{\rm calc}$, of
the Fe K$\alpha$ line given the photon index of the underlying continuum,
the column density $N_{\rm H}$, assumed Fe abundances, and
a covering fraction $f_{\rm c}$ of the sky as seen from the central X-ray source.

With the abundances of Wilms et al. (2001), solar abundance for Fe means
$2.69 \times 10^{-5}$ Fe atoms for every H atom. We see that for the 
material obscuring 4U~1907+09, the Fe abundance is $\sim$4 times this value.
Using the best-fit values of $\Gamma$ (1.15) and $N_{\rm H}$ ($2.18 \times 10^{22}$ cm$^{-2}$)
from the 2007 time-averaged FDCut fit, we find EW$_{\rm calc}$ = 47 eV $\cdot f_{\rm c}$.
With an observed EW of 47$\pm 4$\,eV, 
we can conclude that the Fe K$\alpha$ line is consistent with
being transmitted by the absorbing material if the covering fraction is close to
unity. It is thus likely that the absorbing material along the line of sight
is in the stellar wind in the vicinity of the pulsar.
The high covering factor implied for the Fe line emitter,
the lack of continuum emission observed below
1 keV in the time-averaged spectrum, and the absorbing/Fe-line emitting material being
full-covering are all consistent with a wind accreting geometry,
as opposed to a disk accreting geometry.

The Fe K$\alpha$ line energy, the K$\beta$/K$\alpha$ intensity ratio,
and the properties of the absorber (e.g., edge consistent with neutral Fe K)
all suggest that the bulk of the absorbing/Fe-line transmitting material
is neutral or at most moderately-ionized. 
Theoretical calculations based on Kallman \& McCray (1982), however,  
indicate that the stellar wind should be fully ionized at the radius of the neutron star.
According to Kallman \& McCray iron becomes fully ionized above $log(\xi) = 3$ where 
$\xi$$\equiv L / n R^{2}$.  Using the stellar wind parameters from Cox et al.\ (2005) we 
can calculate the ionization parameter, $log(\xi) = 3.9$, indicating full ionization. 
We calculate the distance from the companion star at which the ionization of the wind is consistent with 
the observed emission features to be roughly 16 times the orbital radius of the pulsar.  
The existence of neutral, line-emitting iron close to the neutron star in the midst of a highly ionized wind 
could be possible if the wind from the companion star is clumpy rather than homogeneous.  
Sako et al. (1999) used the clumpy wind model to reconcile the presence of flourescent lines
in the spectrum of the HMXB Vela X-1 by considering an inhomogeneous wind with
cool, dense clumps of neutral or near-neutral material.  We will show in \S 6.4
using time variability arguments that this model is likely for the accretion in \fu.


\subsection{Variability of the Continuum Emission and Absorbing Components}

Time-resolved spectral modeling shows that
variations in the power-law component are observed in the 1--40 keV continuum variability.
The multi-band lightcurves show correlated flux variations up to 10 keV, in agreement with this notion.
Given the low column density, variations due to the absorbing material cannot account for
continuum variability above 3 keV, particularly in the prolonged dip in the 2006 observation and
the flares in the 2007 observation; though as noted earlier, the change in $\Gamma$ coinciding
with the dip at the end of the 2007 observation may be due to incomplete modelling of a 
partial-covering absorber as indicated by Fig.\ \ref{color07}.

The X-ray continuum in wind-accreting pulsars is generally thought to
originate in the accretion column as the accretion flow is channeled by 
the neutron star's magnetic fields onto its poles and then is decelerated
in a radiative shock and settles onto the surface of the
neutron star. Recent models attribute the X-ray emission
to inverse Comptonization of soft seed photons (originating in the accretion mound and the shock front)
as the photons interact with the compressing gas (e.g., Becker \& Wolff 2007).
Variations in $\Gamma$ could thus indicate changes in the optical depth, temperature, or
geometry and therefore the average number of upscatterings a typical photon undergoes. 

Finally, the observation of a doubling in \NH during the final 
$\sim$30 ks of the 2007 observation further indicates that the absorbing 
material is indeed clumpy.
According to recent calculations, winds of O and B
stars show strong clumping, due to instabilities in the line-driven
acceleration mechanism (Dessart \& Owocki, 2005).  Oskinova et al.\ (2007) 
confirmed this result, showing that clumping is indeed necessary to 
reconcile measured line strengths with observed mass loss rates of the companion star. 
Observationally, in the HMXB Vela X-1, a
system similar to 4U~1907+09, Sako et al. (1999) showed that a clumpy
wind is required to explain the lines seen in the Chandra gratings
spectrum. This result was confirmed in a statistical analysis of the
Vela X-1 flaring behavior (Fuerst et al., 2008),  who found typical
clump masses around 2.3$\times 10^{20}\,$g. 

Flares seen in \fu on the scale of $\sim$1\,hr as in our 2007
observation are consistent with flaring behavior described by Owocki et al. (2009)
and Walter \& Zurita--Heras (2007).  Based on the clumpy wind models, Walter \& 
Zurita--Heras calculated clump masses from the brightness and duration of typical flares.
Their results are on the order of $10^{21}$\,g with \NH values for the clumps
around $10^{22}\,$cm$^{-2}$.  Using similar methods we obtain clump masses 
of around 1.5$\,\times\,10^{20}$\,g and a value for \NH also around $10^{22}\,$cm$^{-2}$.
Additionally, this value for \NH is consistent with the dip in the 2007 observation where
the increase in \NH is about $2 \times 10^{22}\,$cm$^{-2}$ which could be caused by a clump
passing through the line of sight.

From the color-color diagrams we can see that the dip in the 2007 observation and the overall 
dimness of the 2006 observation are consistent with a partial covering absorber in the line of 
sight.  This could be caused by the clumps in the stellar wind partially obscuring the object.  
The prolonged dip in the 2006 observation does not fit this explanation, however it can still 
be explained by a clumpy wind model. Since clumpy winds are expected to have
regions where the wind density is significantly below its average
density this would explain dips due to sudden decreases
in the mass accretion rate as in 2006 and previously observed in \fu by in't Zand et al.\ (1997).
The clumpy wind model therefore explains the observed dipping and flaring behavior
as well as the presence of near-neutral iron in the otherwise highly
ionized stellar wind.


\section{Conclusions}

The use of the \textsl{Suzaku} observatory for our two observations has lead to several advantages over previous
observations of 4U~1907+09 in that the broadband coverage of the XIS+PIN has allowed us to probe in depth the
spectral variability in this source.  The XIS offers CCD-resolution and a high effective area to study Fe K 
bandpass features, plus higher sensitivity below 3\,keV than previous investigations to accurately quantify 
absorption.  The PIN has very low background covering the 13--40\,keV band, allowing us to analyse characteristics 
of the CRSF at 19\,keV.
 
Simultaneous multi-band dips and flares are likely caused by changes in accretion rate possibly 
due to inhomogeneous, clumpy winds.  Dips in the soft band could also be explained by clumpy winds 
when a clump passes through the line of sight.  

We have studied the Fe K $\alpha$ line with unprecedented precision and the Fe K$\beta$ line for the first time.  
We have determined that the iron-line emitting material is very close to the pulsar and therefore could also be 
due to cooler, denser clumps in an otherwise fully ionized stellar wind.

The pulse period has undergone dramatic changes recently and we are able to fill in 
the gap between Fritz et al.\ (2006) and \.Inam et al.\ (2009) where the second torque reversal took place.
The torque reversals cannot be fully explaned by any current model.

Finally, we have observed the CRSF at $\sim$19\,keV in our time-averaged, phase-resolved and time-resolved spectral fitting.
Lack of variation of the cyclotron line energy with flux places \fu in the same regime as A0535$+$262, somewhere in between
the large and small luminosity regimes where perhaps the effect of the ram pressure is just enough to keep the 
shock front of the accretion column at a fairly constant height.


\begin{acknowledgements}
This research has made use of data obtained from the \textsl{Suzaku} 
satellite, a collaborative mission between the space agencies of Japan (JAXA) and the USA (NASA).
This work has made use of HEASARC online services, supported by NASA/GSFC.
The research was supported by NASA contract NAG5--30720, and grants NNX08AC88G and NNX09AG79G.
We acknowledge support from the Bundesministerium f\"ur Wirtschaft und
Technologie under Deutsches Zentrum f\"ur Luft- und Raumfahrt grants
50OR0701, 50OR0808, and 50OR0905. F.\ F\"urst acknowledges support from a
Deutscher Akademischer Austauschdienst stipend.
\end{acknowledgements}


\end{document}